# Approaches for enhancing extrapolability in process-based and data-driven models in hydrology

Haiyang Shi[1,2]

[1] Department of Civil and Environmental Engineering, University of Illinois at Urbana-Champaign, Urbana, IL 61801, USA
[2] State Key Laboratory of Desert and Oasis Ecology, Xinjiang Institute of Ecology and Geography, Chinese Academy of Sciences, Urumqi, 830011, China

Corresponding Author: Haiyang Shi (haiyang@illinois.edu; haiyang.shi9473@outlook.com)

**ABSTRACT**

The application of process-based and data-driven hydrological models is crucial in modern hydrological research, especially for predicting key water cycle variables such as runoff, evapotranspiration (ET), and soil moisture. These models provide a scientific basis for water resource management, flood forecasting, and ecological protection. Process-based models simulate the physical mechanisms of watershed hydrological processes, while data-driven models leverage large datasets and advanced machine learning algorithms. This paper reviewed and compared methods for assessing and enhancing the extrapolability of both model types, discussing their prospects and limitations. Key strategies include the use of leave-one-out cross-validation and similarity-based methods to evaluate model performance in ungauged regions. Deep learning, transfer learning, and domain adaptation techniques are also promising in their potential to improve model predictions in data-sparse and extreme conditions. Interdisciplinary collaboration and continuous algorithmic advancements are also important to strengthen the global applicability and reliability of hydrological models.

# 1 INTRODUCTION

The application of process-based and data-driven hydrological models plays a crucial role in modern hydrological research (Jung et al., 2019; Kraft et al., 2022; Sutanudjaja et al., 2018), particularly in predicting essential water cycle variables such as runoff (Ghiggi et al., 2019; Zhang et al., 2024), evapotranspiration (ET) (Jung et al., 2010; Shi et al., 2022a, 2023; Xie et al., 2021), and soil moisture (Abowarda et al., 2021; Han et al., 2023). These models provide a scientific basis for water resource management, flood forecasting, ecological protection, and optimizing water-energy-food nexus (Cai et al., 2018). Process-based models are widely used for their ability to simulate the physical mechanisms of watershed hydrological processes (Beck et al., 2016; Clark et al., 2015). Concurrently, data-driven models are increasingly applied in hydrological predictions due to the accumulation of data and advancements in machine learning algorithms (Reichstein et al., 2019; Shi et al., 2022a; Xu et al., 2023; Zhang et al., 2024). The integration and comparison of both approaches (Kraft et al., 2022) offer new perspectives for improving the prediction capabilities of the global water cycle.

However, the extrapolation ability of models under different geographical and hydrological conditions (such as drought) remains unclear due to the uneven distribution of current surface hydrological observation stations (e.g., runoff observation stations, flux towers measuring ET) worldwide. In terms of geographical distribution, available surface observation data are predominantly concentrated in economically developed regions such as Europe and North America (Fig. 1). In contrast, data are sparsely distributed in less economically developed and sparsely populated areas (e.g., Africa, the Amazon, Siberia, Central Asia, and the Tibetan Plateau).

These sparsely observed regions play a critical role in the global water cycle due to their vast areas and significant impacts from climate change. Although satellite remote sensing technology significantly contributes to the large-scale estimation of hydrological variables in these regions, many models developed based on remote sensing (e.g., ET estimation and soil moisture estimation using satellite imagery (Han et al., 2023; Mu et al., 2011)) still face limitations due to the calibration and validation of observation data.

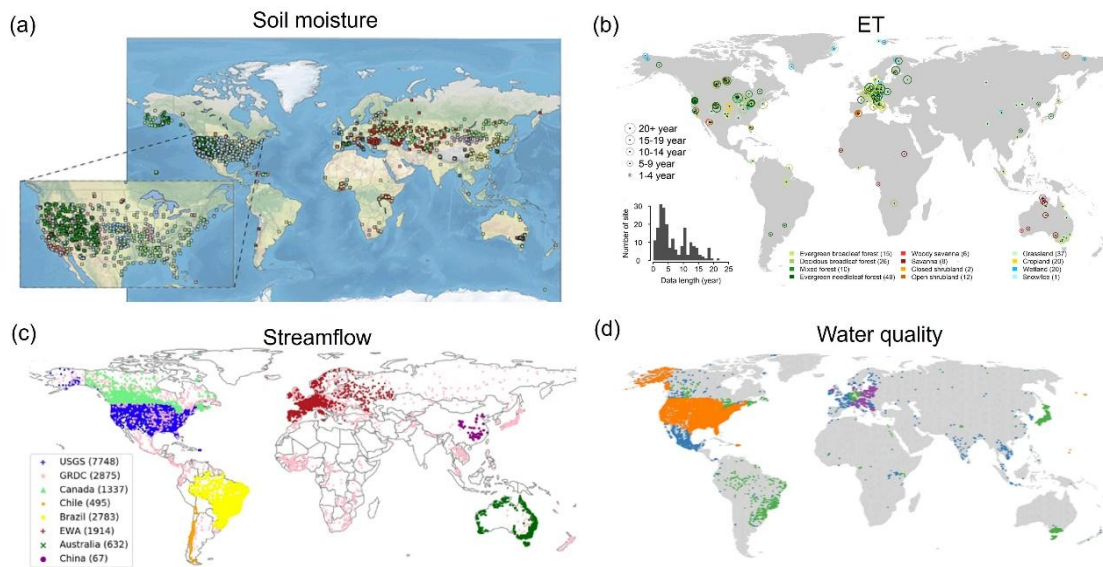

**Figure 1** Distribution of observational data available for calibrating or training global process-based and data-driven hydrological prediction models. (a) Network of soil moisture observation stations, modified from the ref. (Dorigo et al., 2021), (b) Network of flux stations which can monitor evapotranspiration and also carbon flux, modified from the ref. (Pastorello et al., 2020), (c) Network of runoff gauge stations, modified from the ref. (Lin et al., 2019), (d) Network of water quality (total suspended solids) stations, modified from the ref. (Virro et al., 2021).

Process-based models predict hydrological dynamics by simulating physical processes within catchments, such as precipitation, evaporation, infiltration, runoff, and ET (Kraft et al., 2022). These models rely on empirical and semi-empirical representations of physical processes as well as detailed geographical, meteorological, and hydrological data to capture the complex hydrological cycles within a watershed. Although process-based models can perform well in well-gauged catchments, their application in ungauged catchments faces numerous challenges (Guo et al., 2021). Firstly, the significant differences in geographical and meteorological conditions between catchments make it difficult for a model calibrated in one catchment to maintain the same accuracy in another (Guo et al., 2021). Secondly, parameter regionalization (Beck et al., 2016; Guo et al., 2021; Pagliero et al., 2019; Qi et al., 2022) is crucial for enhancing the model's extrapolability, but effectively transferring parameters from gauged to ungauged catchments remains a complex issue (Beck et al., 2016; Lane et al., 2021; Qi et al., 2022).

Data-driven models establish relationships between various hydrological, climatic, soil, and topographical predictors and target variables (such as runoff, ET, and soil moisture) directly from data, and they can effectively simulate and predict in data-rich scenarios (Reichstein et al., 2019). However, the transferability of models trained in data-rich regions to unknown regions remains uncertain, especially when the characteristics of the climate, hydrology, and environment of the target location are not included in the training dataset. Furthermore, even if the training set encompasses the data distribution of the test set, significant prediction errors may still occur. For example, in ET prediction, different plant functional types (PFTs) may have similar predictor values but different ET values and mechanisms (Shi et al., 2023). Moreover, both process-based and data-driven models may perform poorly under extreme hydrological conditions such as extreme runoff events (van Kempen et al., 2021) and severe droughts. Process-based models often exhibit substantial deficiencies in process responses to extreme conditions (e.g., predicting vegetation stomatal conductance during extreme droughts (Damour et al., 2010; Medlyn et al., 2011) in ET estimation). Data-driven methods also suffer from the limited number of observational data under extreme conditions, hindering the learning of relationships under such scenarios.

This paper reviewed and compared current methods for assessing and enhancing the extrapolability of both process-based and data-driven models, discussing their prospects and limitations, and exploring potential future directions. It can provide useful information for future research on global-scale hydrological simulation and prediction.

## 2 EXTRAPOLABILITY DEFINITION AND ASSESSMENT

### 2.1 DEFINITION

The extrapolability (Jung et al., 2020; Meyer and Pebesma, 2021) of hydrological, ecological and geographical models refers to the ability of a model to accurately predict in unobserved regions or under conditions significantly different from those of the training dataset. This concept is of paramount importance for global climate change research, water resource management, and ecological conservation, as many models need to be applied in data-scarce or previously unobserved areas (Alcamo et al., 2003; Sutanudjaja et al., 2018). Specifically, extrapolability includes the following aspects:

(i) Geographical Extrapolability: Geographical extrapolability refers to the ability of a model to make predictions in different geographical locations. This characteristic is crucial for the generalization capability of the model. For instance, a hydrological model trained in European catchments should, if it has good geographical extrapolability, be able to accurately predict runoff and evapotranspiration in North American catchments as well (Zhang et al., 2024). Given that geographical features (such as topography, soil type, vegetation cover, etc.) can vary significantly between regions and watersheds (Wagener et al., 2007), geographical extrapolability requires that the model adapts to these differences and provides accurate predictions.

(ii) Climatic Condition Extrapolability: Climatic condition extrapolability refers to the ability of a model to make accurate predictions under different climatic conditions. Variations in climate, including drought, humidity, cold, or heat, can significantly affect hydrological and ecological processes. For example, a model with good climatic condition extrapolability should perform well under conditions of drought, extreme heat, and extreme humidity. This requires the model not only to capture local climatic characteristics but also to adapt to changes under extreme climatic conditions.

(iii) Long-term temporal Extrapolability: It refers to the ability of a model to make accurate predictions over different long-term periods. This is especially important for long-term climate change research (Hausfather et al., 2020) such as global warming under different future $CO_2$ emission scenarios. For instance, a model trained on historical data, if it has good temporal extrapolability, should be able to provide accurate predictions under future climate scenarios. This requires the model to capture long-term trends and changes and adapt to possible future climatic and environmental changes.

## 2.2 MODEL PERFORMANCE METRICS

To effectively quantify and evaluate the extrapolability of models, researchers have proposed a series of metrics and methods. Prediction error metrics are commonly used evaluation standards, including Root Mean Square Error (RMSE) and Mean Absolute Error (MAE). RMSE is the square root of the average of the squared differences between predicted and actual values, used to measure the accuracy of model predictions. Lower RMSE values indicate better extrapolability of the model. MAE is the average of the absolute differences between predicted and actual values, similar to RMSE, with lower MAE values indicating better model extrapolability. The Nash-Sutcliffe Efficiency (NSE) is also used to evaluate the relative accuracy of model predictions, with values ranging from negative infinity to 1, where 1 indicates a perfect match, 0 indicates the model's predictive performance is equivalent to the mean of the observed data, and negative values indicate the model's performance is worse than the mean. Additionally, the Kling-Gupta Efficiency (KGE) (Knoben et al., 2019) is another commonly used metric. Reliability and robustness metrics are also key in assessing model extrapolability, including extreme value prediction capability and model uncertainty. Extreme value prediction capability evaluates the model's performance in predicting extreme climatic conditions (such as severe droughts or flood events), which can be quantified by comparing the model's prediction errors during extreme events. Model uncertainty is assessed through uncertainty analysis to evaluate the model's stability and reliability under different scenarios. Through these metrics and methods, researchers can systematically quantify and evaluate the extrapolability of hydrological and ecological models, thereby providing the scientific basis for model improvement and application.

## 2.3 EXTRAPOLABILITY ASSESSMENT IN GAUGED AND UNGAUGED CONDITIONS

### 2.3.1 LEAVE-ONE-OUT CROSS-VALIDATION

For a global-scale model, extrapolability needs to be evaluated globally and under various hydrological and climatic conditions. A model that achieves high accuracy only in certain regions or under certain hydrological and climatic conditions cannot be considered to have high extrapolability. Leave-one-out cross-validation (LOOCV) plays a foundational role in the study of global extrapolability of ET and other hydrological models (Jung et al., 2019; Shi et al., 2022a; Zeng et al., 2020; Zhang et al., 2021), particularly through specific applications of leaving out one year and leaving out one site (Fig. 2a), which further refine and evaluate the model's predictive capabilities. The temporal cross-validation method of leaving out one year (Shi et al., 2022a, 2022b) involves using the data from a particular year in the dataset as the test set, while the data from the remaining years are used as the training set for model training and validation. This method effectively evaluates the model's generalization capability in the temporal dimension, especially when assessing the model's prediction accuracy for future climate scenarios or historically unobserved years. By analyzing the model's performance on the left-out year's data, researchers can identify the model's adaptability and stability across different periods, thus optimizing model parameters to enable more accurate predictions of future hydrological and ecological changes. On the other hand, the method of leaving out one site involves using the data from a specific observation site as the test set, while the data from the remaining sites are used as the training set for model training and validation. This method evaluates the model's generalization capability in the spatial dimension, particularly in terms of prediction accuracy under different geographical and climatic conditions. By analyzing the model's performance on the left-out site's data, researchers can identify the model's adaptability in different geographical regions and environmental conditions. The method of leaving out one site is particularly important for evaluating the model's prediction capability in underdeveloped or data-scarce regions, where data are often limited and unevenly distributed. These two LOOCV methods not only reveal performance variations of the model under different training datasets, providing a deep understanding of model stability but also allow for more precise adjustment of model parameters through sequential validation by leaving out one year or one site. This improves the model's prediction accuracy in unobserved regions and under different climatic conditions. Additionally, these methods can be used to assess the model's performance under extreme climatic conditions (such as severe droughts or flood events), analyze the model's prediction errors in these scenarios, and further improve the model to enhance its reliability and accuracy in coping with extreme conditions.

LOOCV can largely be seen as an approximation of a model trained or calibrated using all available data, as it excludes only one site or one year of data each time. It is often considered a baseline evaluation of the extrapolability of a model trained or calibrated using all data. However, it can only be evaluated within the observed data space and cannot be used to assess the model's performance in unknown regions and situations. For example, using LOOCV with FLUXNET2015 sites for ET prediction still makes it difficult to understand the model's accuracy in distant regions such as the Tibetan Plateau, Amazon, Siberia, and Central Asia, and it is challenging to provide a global distribution of the model's expected accuracy estimates.

**2.3.2 GLOBAL MODEL ACCURACY MAPPING BASED ON SIMILARITY**

Global estimates of model accuracy often rely on similarity measures (usually assessed through the distance between training and test set data variables or static attributes) (Jung et al., 2020; Meyer and Pebesma, 2021). First, a method similar to LOOCV is used to obtain the accuracy and similarity or distance of each site relative to the training set. Then, a relationship between the two can be established using regression methods (Jung et al., 2020), thereby achieving accuracy estimates for unknown regions or any unknown test set (Meyer and Pebesma, 2021). The core idea is to calculate the distance between a new prediction location and sample points in the training dataset within the predictor variable space and to use a Dissimilarity Index (DI) (Meyer and Pebesma, 2021) to measure the similarity of these locations. First, the predictor variables are standardized and weighted to reflect their importance in the model. Then, the weighted Euclidean distance between the new prediction location and training data points is calculated, and the closest training point is identified. The DI is calculated by dividing the distance between the new location and the nearest several training points (Fig. 2b) by the average distance between training data points. This index defines the model's area of applicability, ensuring that prediction errors within these areas are comparable to cross-validation errors in the training data. By quantifying the similarity between new data points and training data, this method improves the model's predictive reliability and generalization ability in unknown regions. Similar approaches have been used to evaluate the extrapolability of carbon flux models (Jung et al., 2020) and global runoff prediction models (Nearing et al., 2024). However, despite the ability to measure similarity using various distance metrics, accurately mapping 'accuracy distance' remains a challenge (Nearing et al., 2024; Shi et al., 2023). This may be related to the uncertainty in the weights of distances for various predictor variables. For instance, if certain variables at some sites are not similar to those in the training set but do not play an important role in the model's predictions, this dissimilarity may not necessarily lead to low model accuracy.

# 3 APPROACHES FOR ENHANCING EXTRAPOLABILITY

## 3.1 USING MODEL PARAMETERS AND TRAINING DATA BASED ON SIMILARITY

In the process of hydrological model parameterization, methods based on hydrological similarity and watershed classification (Ciulla and Varadharajan, 2024; Wagener et al., 2007) can significantly enhance the predictive capability and generalization performance of models (Beck et al., 2016; Guo et al., 2021; Song et al., 2022). Firstly, by defining similarity metrics (such as Euclidean distance and cosine similarity), the similarity between two catchments across multiple hydrological variables can be quantified (Guo et al., 2021). These metrics are used to identify and quantify the similarity between different catchments, thereby providing a basis for the transfer and sharing of model parameters.

Multivariate similarity analysis is key to this method. By comprehensively considering multiple hydrological and meteorological variables (such as precipitation, temperature, and evapotranspiration) and static properties, the similarity between catchments can be assessed(Ciulla and Varadharajan, 2024). Techniques like Principal Component Analysis (PCA) can be used to identify and utilize the primary similarity features in high-dimensional data, helping to determine which catchments can be referenced for parameterization. For example, PCA can reduce multiple variables to a few principal components that explain most of the variance in the data, simplifying the complexity of similarity analysis. Based on similarity analysis, parameter transfer is an effective method. By transferring parameters from a catchment with observational data to a similar ungauged catchment, the model's prediction accuracy can be significantly improved. This method assumes that similar catchments exhibit similar hydrological responses, allowing them to share model parameters. For instance, previous research (Beck et al., 2016) considered the similarity of factors such as topography, soil type, vegetation cover, and climatic conditions to match the global catchments with the calibrated parameters of the 10 most similar 'donor' catchments (Fig. 2c). It showed that using parameters provided by 'donors' can achieve higher accuracy in most catchments, thus improving the model's global generalization ability and extrapolability.

In data-driven models, using this pre-screening method based on similarity for data selection can also improve accuracy. For example, it was found that in ET simulation, using a subset of data from specific PFTs (Plant Functional Types) rather than all data can achieve significantly higher accuracy at some sites such as in wetlands (Shi et al., 2023). This method can reduce confusion between different PFTs to some extent. However, when the data richness within the subset (which has high internal similarity) is not very high, the effectiveness of this method may be limited, as it may arbitrarily reduce the model's ability to learn useful relationships between variables from other PFTs.

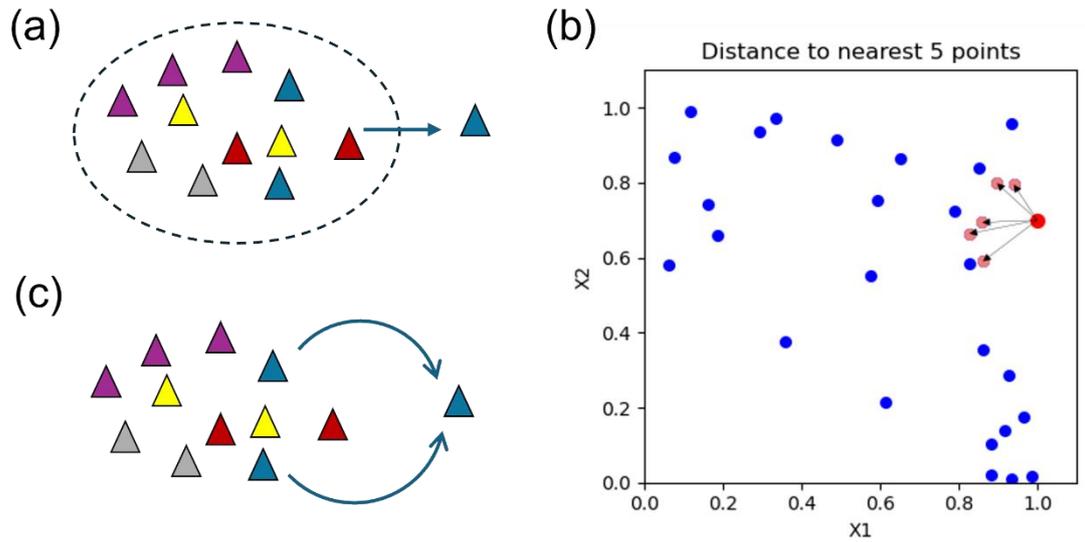

**Figure 2** Extrapolability assessment using LOOCV (a) and Dissimilarity Index calculated by the distance to nearest points in predictor space (b) and enhancing extrapolability by using similarity-based model parameters (c).

## 3.2 DEEP LEARNING AND DOMAIN ADAPTATION

Deep learning can significantly improve the extrapolability of hydrological prediction models through various methods (Feng et al., 2021; Kratzert et al., 2019; Zhang et al., 2024). Firstly, deep learning models can handle and learn from large, diverse datasets, thus enhancing their generalization ability (Reichstein et al., 2019). By training models using global datasets from different geographical and climatic conditions, the models can learn more comprehensive features and patterns, thereby improving prediction accuracy in ungauged regions. Additionally, deep learning models using encoder-decoder architectures, such as Encoder-Decoder Double-Layer Long Short-Term Memory (ED-DLSTM) (Zhang et al., 2024), can effectively capture complex spatiotemporal dependencies in time series data. This architecture not only considers the temporal dependencies of the input data but also integrates static spatial attributes (such as topography, soil type, and vegetation cover) and temporal forcing attributes (such as precipitation and temperature), thereby enhancing prediction performance. Furthermore, deep learning models achieve adaptation in different regions and conditions through transfer learning and domain adaptation techniques. Transfer learning involves training the model in a source domain (a region with abundant data) and then fine-tuning it in the target domain (a region with scarce and sparse data), thus improving the model's predictive performance in the data-sparse regions (Xu et al., 2023).

Domain adaptation (Ben-David et al., 2006; Farahani et al., 2021; Ma et al., 2018) reduces the performance decrease between different domains by adjusting the distribution differences between training and test data (Fig. 3a). For instance, feature alignment methods adjust the feature distribution between training and test data, ensuring consistent input feature spaces across different domains. Common techniques include Maximum Mean Discrepancy (MMD) and adversarial training. For example, using Generative Adversarial Networks (GANs) (Perera et al., 2024) and Domain-Adversarial Neural Networks (DANNs) (Ganin et al., 2016; Ma et al., 2021) can generate data similar to the target domain distribution, thereby improving the model's adaptability and generalization ability in the target domain (Fig. 3b). Additionally, deep learning models can assimilate various types of data, including ground observations and remote sensing data, thereby compensating for the limitations of single data sources. For example, integrating satellite-observed soil moisture information or flow duration curves can enhance the model's predictive capability in data-scarce regions (Guo et al., 2021). Furthermore, through ensemble techniques, deep learning models can further improve prediction accuracy and stability. Different model configurations can capture various features and patterns in the data, and ensemble models combine the predictions of multiple models, reducing the errors and uncertainties of individual models, and thus enhancing the model's generalization ability.

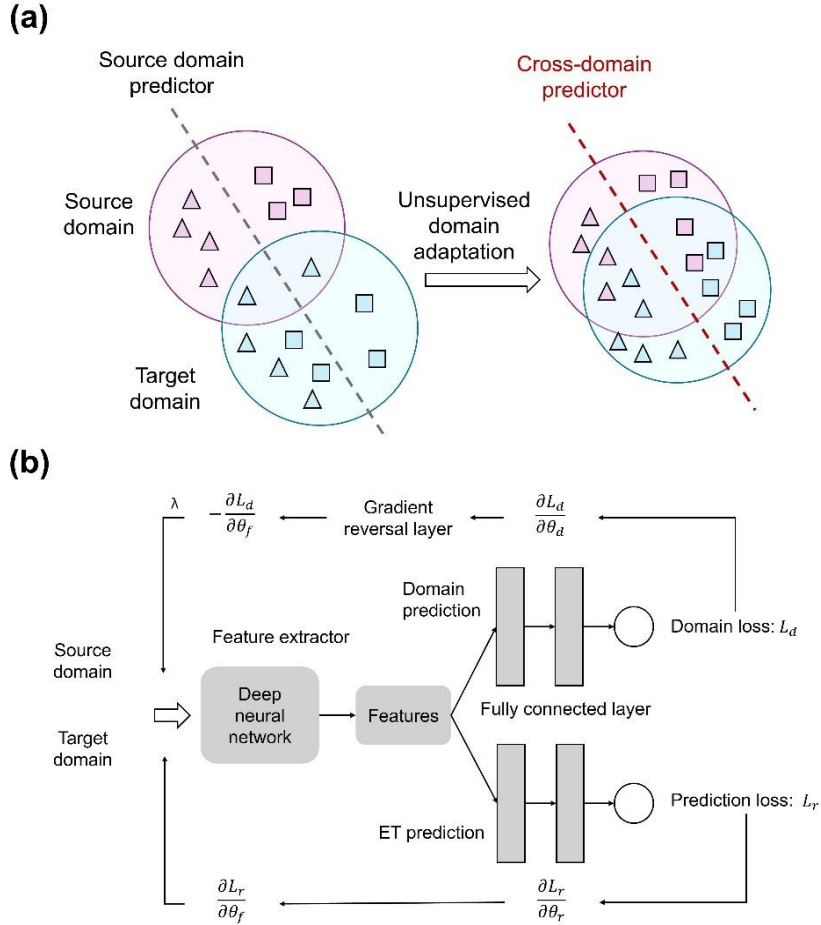

**Figure 3** (a) The concept of unsupervised domain adaptation. Before applying unsupervised domain adaptation, the predictor trained on the source domain tends to make inaccurate predictions in the target domain. However, after implementing unsupervised domain adaptation, the source and target domains become aligned, enabling the cross-domain predictor to make accurate predictions in both domains. (b) The architecture of the DANN for ET prediction. $\theta_f$, $\theta_d$, and $\theta_r$ denote trainable model weights in the feature extractor, domain classifier (for domain prediction), and ET predictor (for ET prediction), respectively.

$L_r$ denotes the regression loss and $L_d$ denotes the domain loss.

## 4 CHALLENGES AND FUTURE DIRECTIONS

### 4.1 MORE ACCURATE CATCHMENT CLASSIFICATION AND SIMILARITY ASSESSMENT

Classifying catchments with similar biophysical characteristics helps to understand and predict their hydrological behaviour. The potential of this method lies in its ability to utilize diverse datasets that include detailed information about land surface structure, climatic conditions, vegetation cover, land use, and human impact (Ciulla and Varadharajan, 2024; Wagener et al., 2007). This data-driven approach enables precise catchment classification on both regional and global scales, supporting more refined hydrological modelling and management. By using machine learning algorithms (such as k-means and hierarchical clustering), researchers can effectively classify catchments into different groups, thereby identifying catchments with similar hydrological behaviour under different geographical and climatic conditions, and providing support for hydrological prediction in ungauged catchments.

In evaluating and enhancing the extrapolability of models through catchment classification methods, accurate classification methods can significantly improve the predictive accuracy of models in different regions. However, inaccurate classification may lead to uncertainty in model parameter transfer, thereby affecting the overall performance of the model. Firstly, the classification system groups catchments by establishing a baseline of similarities and differences between them. This method uses feature vectors and distance metrics in a multidimensional space to ensure the scientific accuracy of the classification results. However, the issue of multicollinearity, where multiple variables provide similar information, can lead to inaccurate classification results (Ciulla and Varadharajan, 2024). Multicollinearity makes it difficult for the classification model to distinguish which variables are important, thus affecting the accuracy of parameter transfer. For example, if two catchments have different hydrological responses in reality but are incorrectly classified as similar due to multicollinearity in the high-dimensional feature space, the parameters transferred from one catchment to the other may not be applicable, leading to increased prediction errors. Moreover, the problem of dimensionality disaster is another major cause of inaccurate classification (Ciulla and Varadharajan, 2024). As the number of features used for classification increases, the density of data points in the feature space significantly decreases, making the feature space sparse and increasing computational complexity. This 'dimensionality disaster' makes distance-based metrics unreliable in high-dimensional space, thus affecting the accuracy of classification results. For instance, when catchments are incorrectly classified into different groups, the parameters transferred from the wrongly classified group may not be suitable for the target catchment, resulting in inaccurate predictions. The heterogeneity between catchments in terms of their topography, soil, vegetation cover, and climatic conditions is significant. If the classification method fails to accurately capture these differences and instead bases the classification on insufficient or misleading information, these differences will cause significant model uncertainty during parameter transfer.

This primarily affects model construction by reducing prediction accuracy, increasing model uncertainty, and impacting model reliability. Therefore, accurate classification methods are crucial in the process of catchment classification and model parameter transfer. By overcoming issues of multicollinearity, dimensionality disaster, and catchment feature heterogeneity, the accuracy of classification can be improved, ensuring the effectiveness of parameter transfer and the extrapolability of the model. This will provide more reliable scientific support for global hydrological prediction and management while also demonstrating the great potential of classification methods in the development and application of hydrological models.

**4.2 FOCUS ON REMOTE OR EXTREME CONDITIONS AND ADVANCED ALGORITHMS**

Current methods for evaluating model extrapolability need to pay more attention to remote regions and extreme climatic conditions, as these areas often lack observational data but are of significant importance (Joetzjer et al., 2013) in the context of global climate change. In remote regions where ground observation data is scarce, remote sensing data can provide essential supplementation. For example, satellite observation data can provide information on such as soil moisture and it can be used to improve model prediction accuracy in these areas through data assimilation methods (Azimi et al., 2020; Wakigari and Leconte, 2023). Multi-source remote sensing data can also help assess the multidimensional similarity between these remote areas and observed regions. By integrating multi-scale models, different scale hydrological processes can be better captured. For instance, regional climate models and global climate models can be combined to provide more detailed and accurate climate predictions. This multi-scale integration method can improve the measurement of similarity at both large and small scales. Existing models often lack predictive capability under extreme climatic conditions. Enhancing simulations of extreme events, such as floods and droughts, can improve model prediction accuracy under extreme climatic conditions.

This requires adding data on extreme climate events to the training dataset to ensure that the model can learn and adapt to these events.

As computing power and data acquisition methods continue to improve, the application of deep learning, transfer learning, and domain adaptation algorithms in global hydrological prediction will become more widespread. These technologies can provide high-precision predictions, particularly in data-scarce regions and under extreme climatic conditions, thus improving the efficiency and scientific basis of global water resource management and disaster response. In the future, interdisciplinary collaboration will further drive the development and application of these technologies. Collaboration among hydrologists, climatologists, data scientists, and computer scientists can lead to more innovative solutions, enhancing the predictive capabilities and application breadth of models. With continuous algorithm improvements and the emergence of new technologies, these advanced methods will make greater progress in improving the extrapolability of hydrological predictions. For example, combining graph neural networks (GNNs) (Sun et al., 2021) and spatiotemporal transformer models (Rasiya Koya and Roy, 2024; Xu et al., 2023) can further enhance the model's adaptability to complex environments.

## 5 CONCLUSION

The integration of process-based and data-driven hydrological models plays a pivotal role in modern hydrological research by providing essential predictions for water cycle variables such as runoff, evapotranspiration, and soil moisture. Both approaches offer unique strengths and face specific challenges, particularly in terms of their extrapolation capabilities in diverse geographical and hydrological conditions. Process-based models excel in simulating physical hydrological processes but encounter difficulties in ungauged catchments due to significant geographical and meteorological variations. On the other hand, data-driven models leverage large datasets and advanced machine learning algorithms to predict hydrological dynamics but struggle with transferability to unknown regions, especially under extreme conditions. Accurate assessment and enhancement of model extrapolability are crucial for global-scale applications. Leave-one-out cross-validation (LOOCV) and similarity-based methods provide valuable insights into model performance in unobserved regions and under varying conditions. However, the reliance on similarity metrics and the challenge of accurately mapping 'accuracy-distance' relationships highlight the need for more sophisticated techniques. Deep learning, transfer learning, and domain adaptation algorithms offer promising solutions to these challenges, enabling better handling of diverse and extreme conditions. Future research should focus on improving catchment classification and similarity assessment methods, addressing issues like multicollinearity and dimensionality disaster, and enhancing the reliability and accuracy of parameter transfer. Additionally, the application of multi-scale models and the integration of remote sensing data will further improve model predictions in remote and data-scarce regions. Interdisciplinary collaboration and continuous advancements in computational power and algorithms will drive significant progress in hydrological prediction, ensuring more robust and reliable models for global water resource management and climate change adaptation.

## CONFLICT OF INTEREST STATEMENT

There are no conflicts of interest among the authors.

## ACKNOWLEDGMENTS

This research has been supported by National Natural Science Foundation of China (grant no. U1803243).

## DATA AVAILABILITY

No data is used in this paper.

Sciences 21, 961–976. https://doi.org/10.5194/nhess-21-961-2021

Virro, H., Amatulli, G., Kmoch, A., Shen, L., Uuemaa, E., 2021. GRQA: Global River Water Quality Archive. Earth System Science Data 13, 5483–5507. https://doi.org/10.5194/essd-13-5483-2021

Wagener, T., Sivapalan, M., Troch, P., Woods, R., 2007. Catchment Classification and Hydrologic Similarity. Geography Compass 1, 901–931. https://doi.org/10.1111/j.1749-8198.2007.00039.x

Wakigari, S.A., Leconte, R., 2023. Exploring the utility of the downscaled SMAP soil moisture products in improving streamflow simulation. Journal of Hydrology: Regional Studies 47, 101380. https://doi.org/10.1016/j.ejrh.2023.101380

Xie, M., Luo, G., Hellwich, O., Frankl, A., Zhang, W., Chen, C., Zhang, C., De Maeyer, P., 2021. Simulation of site-scale water fluxes in desert and natural oasis ecosystems of the arid region in Northwest China. Hydrological Processes 35, e14444. https://doi.org/10.1002/hyp.14444

Xu, Y., Lin, K., Hu, C., Wang, S., Wu, Q., Zhang, L., Ran, G., 2023. Deep transfer learning based on transformer for flood forecasting in data-sparse basins. Journal of Hydrology 625, 129956. https://doi.org/10.1016/j.jhydrol.2023.129956

Zeng, J., Matsunaga, T., Tan, Z.-H., Saigusa, N., Shirai, T., Tang, Y., Peng, S., Fukuda, Y., 2020. Global terrestrial carbon fluxes of 1999–2019 estimated by upscaling eddy covariance data with a random forest. Scientific Data 7. https://doi.org/10.1038/s41597-020-00653-5

Zhang, B., Ouyang, C., Cui, P., Xu, Q., Wang, D., Zhang, F., Li, Z., Fan, L., Lovati, M., Liu, Y., Zhang, Q., 2024. Deep learning for cross-region streamflow and flood forecasting at a
31